\documentclass{aa} % for a referee version
\usepackage[colorlinks=true, linkcolor=blue, urlcolor=blue, citecolor=blue, breaklinks=true]{hyperref}      %% to avoid \citeads line fills, add "draft" 
\usepackage[table]{xcolor}
\usepackage{colortbl}
\usepackage{orcidlink}
\setcitestyle{aa}
\usepackage{graphicx}
\usepackage{txfonts}
\usepackage{upgreek}
\def    \apjl           {\rm {ApJL}}

\def    \apjl           {\rm {ApJL}}

\def    \cm             {\,{\rm {cm}}}
\def    \mm             {\,{\rm mm}}
\def    \K              {\,{\rm {K}}}
\def    \g              {\,{\rm {g}}}
\def    \mum            {\,{\upmu \rm{m}}}

\def \bea {\begin{eqnarray}}
\def \ena {\end{eqnarray}}

\def    \C      {{\rm C}}

\def    \cm     {\,{\rm cm}}

\def    \g      {\,{\rm g}}

\def    \H      {{\rm H}}

\def    \pc     {\,{\rm pc}}

\def    \au     {\,{\rm au}}

\def    \kyr    {\,{\rm kyr}}

\begin{document}

   \title{The Rosetta Stone Project}

%   \subtitle{II. On the $L/M$ indicator for the evolutionary stages of star-forming clumps}
   \subtitle{II. The correlation between star formation efficiency and $L/M$ indicator for the evolutionary stages of star-forming clumps in post-processed radiative magnetohydrodynamics simulations}

   \authorrunning{N.-D. Tung et al.}
   \titlerunning{The Rosetta Stone Project. Paper II}

   \author{Ngo-Duy Tung \inst{\ref{CEA}}\fnmsep\thanks{\email{duy-tung.ngo@cea.fr}}\orcidlink{0009-0009-8545-2682}
          \and
          Alessio Traficante \inst{\ref{IAPS}}\orcidlink{0000-0003-1665-6402}
          \and
          Ugo Lebreuilly \inst{\ref{CEA}}\orcidlink{0000-0001-8060-1890}
          \and
          Alice Nucara \inst{\ref{IAPS},\ref{ToV}}\orcidlink{0009-0005-9192-5491}
          \and
          Leonardo Testi \inst{\ref{DIFA},\ref{Arcetri}}\orcidlink{0000-0003-1859-3070}
          \and
          Patrick Hennebelle \inst{\ref{CEA}}\orcidlink{0000-0002-0472-7202}
          \and
          Ralf S. Klessen\inst{\ref{Heidelberg1},\ref{Heidelberg2},\ref{Cfa},\ref{Radcliffe}}\orcidlink{0000-0002-0560-3172}
          \and
          Sergio Molinari \inst{\ref{IAPS}}\orcidlink{0000-0002-9826-7525}
          \and
          Veli-Matti Pelkonen \inst{\ref{IAPS}}\orcidlink{0000-0002-8898-1047}
          \and
          Milena Benedettini \inst{\ref{IAPS}}\orcidlink{0000-0002-3597-7263}
          \and
          Alessandro Coletta \inst{\ref{IAPS}}\orcidlink{0000-0001-8239-8304}
          \and
          Davide Elia\inst{\ref{IAPS}}\orcidlink{0000-0002-9120-5890}
          \and
          Gary A. Fuller \inst{\ref{JBCA},\ref{Koln}}\orcidlink{0000-0001-8509-1818}
          \and
          Stefania Pezzuto\inst{\ref{IAPS}}\orcidlink{0000-0001-7852-1971}
          \and
          Juan D. Soler\inst{\ref{IAPS}}\orcidlink{0000-0002-0294-4465}
          \and
          Claudia Toci\inst{\ref{Arcetri},\ref{ESO}}\orcidlink{0000-0002-6958-4986}
          }

   \institute{Université Paris-Saclay, Université Paris Cité, CEA, CNRS, AIM, 91191, Gif-sur-Yvette, France\label{CEA}
         \and
             INAF -- Istituto di Astrofisica e Planetologia Spaziali (INAF-IAPS), Via Fosso del Cavaliere 100, I-00133, Roma, Italy\label{IAPS}
         \and
             Alma Mater Studiorum Università di Bologna, Dipartimento di Fisica e Astronomia (DIFA), Via Gobetti 93/2, I-40129, Bologna, Italy\label{DIFA}
         \and
             INAF -- Osservatorio Astrofisico di Arcetri, Largo E. Fermi 5, I-50125, Firenze, Italy\label{Arcetri}
         \and
             Dipartimento di Fisica, Università di Roma Tor Vergata, Via della Ricerca Scientifica 1, I-00133 Roma, Italy\label{ToV}
        \and
            Universität Heidelberg, Zentrum für Astronomie, Institut für Theoretische Astrophysik, Albert-Ueberle-Str. 2, 69120 Heidelberg, Germany\label{Heidelberg1}
        \and
            Universität Heidelberg, Interdisziplinäres Zentrum für Wissenschaftliches Rechnen, Im Neuenheimer Feld 205, 69120 Heidelberg, Germany \label{Heidelberg2}
        \and
            Harvard-Smithsonian Center for Astrophysics, 60 Garden Street, Cambridge, MA 02138, U.S.A. \label{Cfa}
        \and
            Radcliffe Institute for Advanced Studies at Harvard University, 10 Garden Street, Cambridge, MA 02138, U.S.A. \label{Radcliffe}
        \and
            Jodrell Bank Centre for Astrophysics, Department of Physics \& Astronomy, The University of Manchester, Oxford Road, Manchester M13 9PL, UK \label{JBCA}
        \and
            Physikalisches Institut der Universität zu Köln, Zülpicher Str. 77, D-50937 Köln, Germany \label{Koln}
        \and
            European Southern Observatory (ESO), Karl-Schwarzschild-Strasse 2, 85748, Garching bei M\"{u}nchen, Germany\label{ESO}        
             }

   \date{Received 26 March 2025; accepted 3 July 2025}

% \abstract{}{}{}{}{} 
% 5 {} token are mandatory
 
  \abstract
   {The evolution of massive star-forming clumps that are progenitors of high-mass young stellar objects are often classified based on  a variety of observational indicators ranging from near-infrared to radio wavelengths. Among them, the ratio of the bolometric luminosity to the mass of their envelope, $L/M$, has been observationally diagnosed as a good indicator for the evolutionary classification of parsec-scale star-forming clumps in the Galaxy.}
   {We   developed the Rosetta Stone project—an end-to-end framework designed to enable an accurate comparison between simulations and observations for investigating the formation and evolution of massive clumps. In this study, we calibrate the $L/M$ indicator in relation to the star formation efficiency (SFE) and the clump age, as derived from our suite of simulations.}
   {We performed multi-wavelength radiative transfer post-processing of radiative magnetohydrodynamics (RMHD) simulations of the collapse of star-forming clumps fragmenting into protostars. We generated synthetic observations to obtain far-infrared emission from $70$ to $500\mum$, as was done in the Hi-GAL survey, and at $24\mum$ in the MIPSGAL survey, which were then used to build the spectral energy distributions (SEDs) and estimate the $L/M$ parameter. An additional $1.3\mm$ wavelength in ALMA Band 6 was also  produced for the comparison with observational data. We  applied observational techniques—commonly employed by observers—to the synthetic data in order to derive the corresponding physical parameters.}
   {We find a correlation between $L/M$ and the SFE, with a power-law form $L/M\propto {\rm SFE}^{1.20^{+0.02}_{-0.02}}$. This correlation is independent of the mass of the clumps and the choice of initial conditions of the simulations in which they formed. The relation between $L/M$ and the ages of the clumps is instead mass-dependent, and can also be strongly influenced by the intensity of the magnetic fields.} 
  {Our results suggest that $L/M$ is a reliable parameter for characterizing the overall evolutionary stage of a given star-forming region. Its value can be directly compared with the star formation efficiency (SFE) parameter derived from simulations. However, to accurately infer the age of the observed clumps, it is essential to constrain their mass.}

   \keywords{methods: numerical --
                methods: statistical --
                star: formation --
                stars: massive --
                ISM: structure
               }

   \maketitle
%
%-------------------------------------------------------------------

%%%%%%%%%%%%%%%%%%%%%%%%%%%%%%%%%%%%%%%%%%%%%%%%%%%%%%%%%%%%%%%%%%%%%%%%%%%%%%%%%%%%%%%%%%%%%%%%%%%%%%%%%%%%%%%%%%%%%%%%%%%%%%%%%%
\section{Introduction}
Dense and massive parsec-scale clumps embedded in giant molecular clouds are known to be the formation sites of intermediate- and high-mass stellar clusters \citep{Motte18b}. Early models of high-mass star formation had predicted a different role of clumps in regulating the seeds for the cluster formation. On the one hand, in what is known as the core-fed scenario, self-gravitating pressure-confined cores form individual massive stars within the fragmenting clumps (e.g., \citealt{2003ApJ...585..850M}). On the other hand,  in the clump-fed model, the initial seeds  simultaneously form massive stars and star clusters, and they draw material in a dynamically complex fashion from the parent clumps (e.g., \citealt{2001MNRAS.323..785B}, \citealt{Vazquez-Semadeni19}, \citealt{Vazquez-Semadeni25}). The debate is now moving beyond this dichotomy in light of recent observational studies, where the vast majority of the clumps support the clump-fed scenario (e.g., \citealt{2019ApJ...886...36S, 2022A&A...662A...8M, 2022A&A...664A..26P, 2023MNRAS.520.2306T, 2024ApJ...966..171M, 2024ApJS..270....9X, 2025A&A...696A.151C}, although a handful of massive regions show properties of fragmentation in agreement with a core-fed scenario; e.g., \citealt{Nony18}, \citealt{Valeille-Manet25}). 

Investigating the evolutionary phases and lifetimes of high-mass star-forming objects is a key objective; it is essential, for example, for characterizing the impact of protostellar accretion luminosity on cluster formation (\citealt{2011ApJ...731...90D,2014ApJ...786..116B,2015ApJ...815..130G, 2016ApJ...822...59S, 2018A&A...615A..94F, 2019MNRAS.483.5355M}). Among the observational diagnostics for the evolutionary classification of the progenitors of clumps, the bolometric luminosity-to-mass of the envelope ratio ($L/M$) has been shown to be a robust and reliable evolutionary indicator (\citealt{2008A&A...481..345M,2013A&A...558A.125D,2016ApJ...826L...8M,2020MNRAS.496.3482P}), along with the ratio of submillimetric luminosity to bolometric luminosity ($L_{\rm submm}/L_{\rm bol}$) and the bolometric temperature (see, e.g., \citealt{2016MNRAS.461.1328E}). In particular, \cite{2016ApJ...826L...8M} used SEPIA/APEX observations of $\C\H_3\C_2\H$($J=12-11$) to calibrate $L/M$ and found three ranges of $L/M$ that they associate with three phases of star formation in massive clumps: 1) a quiescent phase, nondetectable in $\C\H_3\C_2\H$ ($L/M \lesssim 1$), of ongoing formation of relatively low-mass objects; 2) the intermediate phase ($1 \lesssim L/M \lesssim 10$) in which star formation activities within the clump provide enough energy to warm up its envelope and the average clump temperature rises up to ${\simeq}30\K$; and 3) a phase ($L/M \gtrsim 10$) where the clump starts to be warmed up by a drastic increase in luminosity and temperature, and the most massive clumps show the formation of young \textsc{H ii} regions \citep{Cesaroni15}.

The $L/M$ parameter is suggested to be a good indicator of  clump evolution largely based on observational data. To better understand the underlying physical reason, however, requires well-tuned simulations of clump formation and evolution. The comparison between observations and model allows us to quantify how $L/M$ relates to the evolutionary stage of the clump, measured by parameters such as the clump age or the mass already locked into stars,  called star formation efficiency (SFE). The SFE parameter in particular can be quantified in simulations by comparing the total mass in gas with the mass locked into sink particles. This quantity is also independent of the initial clump mass and does not require us to define a zero-age time for clump formation. 

To compare $L/M$ as derived form observations with the SFE extracted from simulations, we opted for an approach combining radiative transfer modeling and synthetic observations of protostellar collapse simulations, as has been done  for molecular clouds (e.g., \citealt{2025A&A...695A..77D,2023ApJ...950...88J,2017ApJ...849....1K,2017ApJ...849....2K}). In particular, we performed detailed radiative transfer post-processing and synthetic observations for the Herschel Infrared Galactic Plane Survey (Hi-GAL; \citealt{2016A&A...591A.149M}) carried out in five infrared (IR) continuum bands between $70$ and $500\mum$, the Spitzer MIPSGAL survey (\citealt{2009PASP..121...76C}) at $24\mum$, and the ALMA SQUALO survey (\citealt{2023MNRAS.520.2306T}) at $1.3\mm$, the last of which will be analyzed in detail in Nucara et al. (2025, Paper III). We then used the Hybrid photometry and extraction routine (\textsc{Hyper}; \citealt{2015A&A...574A.119T}) to extract the clumps from the synthetic maps and computed the bolometric luminosities and the temperatures from the spectral energy distributions (SEDs) of the extracted sources in a similar fashion to what is found in the literature (\citealt{2015MNRAS.451.3089T, Paulson18, Traficante20, 2025A&A...696A.151C}). The resulting values of $L/M$  are analyzed in relation to the direct evolutionary parameters such as the physical time and the SFE in the simulations to provide a semi-empirical calibration between the observationally accessible $L/M$ indicator and the inaccessible lifetimes and SFE of the clumps.

This study is part of the Rosetta Stone (RS) project,\footnote{\url{https://www.the-rosetta-stone-project.eu}} developed under the ECOGAL synergy grant (PIs: P. Hennebelle, S. Molinari, L. Testi, and R. Klessen). Our goal is to understand the formation of massive clumps with a comprehensive analysis of the various mechanisms that participate in the formation of the final protostellar cluster, and consists of three steps: 1) the development of a suite of state-of-the-art radiative magnetohydrodynamics (RMHD) simulations of protostellar collapse and clump formation within molecular clouds under different initial conditions such as clump mass, clump initial turbulence level, or intensity of magnetic fields (called Rosetta Stone 1.0, RS1.0, Lebreuilly et al. 2025, hereafter Paper I); 2) the radiative transfer post-processing and production of synthetic observations of these simulations at various wavelengths in the mid-IR, far-IR, sub-millimeter, and millimeter regimes (this work); and 3) a final processing to mimic real observations and a direct comparisons with them in the most realistic way, particularly tuned for recent ALMA surveys (Nucara et al. 2025, hereafter Paper III). 

The paper is organized as follows. In Section \ref{sec:model_methodology} we present the suite of simulations used in this work and the methodology chosen to produce our synthetic observations in the mid-IR, far-IR, and millimeter regime. In Section \ref{sec:analysis} we derive the $L/M$ parameter mimicking the observational approach, and we compare it with the SFE and clump age. In Section \ref{sec:conclusions} we derive our conclusions.

%%%%%%%%%%%%%%%%%%%%%%%%%%%%%%%%%%%%%%%%%%%%%%%%%%%%%%%%%%%%%%%%%%%%%%%%%%%%%%%%%%%%%%%%%%%%%%%%%%%%%%%%%%%%%%%%%%%%%%%%%%%%%%%%%%
\section{Synthetic modeling methodology}\label{sec:model_methodology}

\subsection{Numerical simulations of cloud collapse and clump formation}\label{sec:sims}

We first briefly present the numerical simulations that are used in this study (for a detailed description of the numerical setups of the simulations, we refer to Paper I).

The suite of simulations is performed with the adaptive mesh refinement (AMR, \citealt{1984JCoPh..53..484B}) code \textsc{Ramses} (\citealt{2002A&A...385..337T}; \citealt{2006A&A...457..371F}) that solves the MHD equations in the ideal MHD regime using the finite volume method across a computational domain of ${\approx}1.53\pc$ in length to follow the dynamical evolution of the clumps up to a maximum resolution of $\Delta x_{\rm max} \sim 38\au$. As representative of the clumps observed in the Hi-GAL and SQUALO surveys, two values of the initial gas mass are considered, $M_{\rm cld} = 500, 1000\,M_{\odot}$, along with one clump radius, $R_{0} \sim 0.38\pc$. A uniform temperature of $10\K$ and a uniform density of $3\times10^{-19}\g\cm^{-3}$ is assumed for the initial state of the gas. At the start of the simulations, the clump is initialized with supersonic turbulent velocities at $\mathcal{M} = 7,10$ with random phases and a Kolmogorov power spectrum of $k^{-11/3}$ (\citealt{1941DoSSR..30..301K}). Two different turbulent seeds (numbered 1 and 2) are investigated to account for stochastic fluctuations. The magnetic field, initially aligned along the $z$-direction, is set according to three values of the mass-to-flux to critical-mass-to-flux ratio which represent three relative potential magnetization scenarios: a quasi-hydrodynamical $\mu=100$, a moderately magnetized $\mu=10$, and a strongly magnetized one $\mu=3$. Sink particles (\citealt{2014MNRAS.445.4015B}), which form when the density reaches $n_{\rm thre} = 10^{9} \cm^{-3}$, are used as sub-grid models to account for the presence of fully formed stars. 

\begin{figure}
    \centering
    \includegraphics[width=0.45\textwidth]{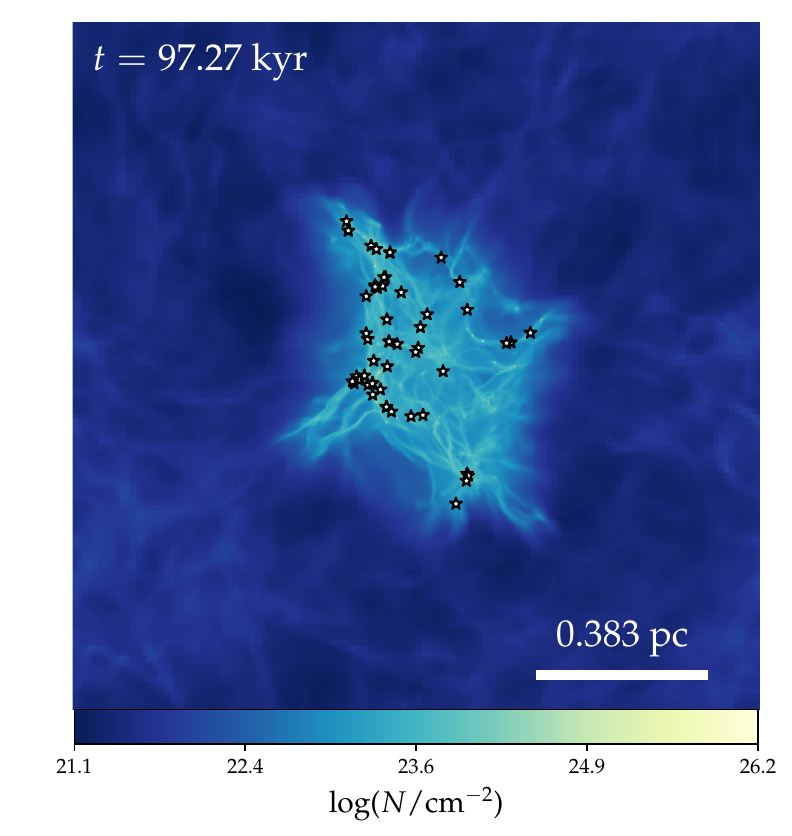}
    \caption{Column density map integrated along the $z$-direction of one of the models used for this study, seen at a full simulation box's length. The star symbols indicate the sink particle positions.}
    \label{fig:coldens}
\end{figure}

An example of the results of the simulations is in Figure \ref{fig:coldens}. Here, we show the column density map of the simulated box along the $z$-direction at an output corresponding to an initial gas mass of $M_{0} = 1000\,M_{\odot}, R_{0} \sim 0.38\pc, \mu=100, \mathcal{M} = 7$. We refer hereafter to this simulation model as the fiducial model.

For the purposes of this work we analyzed in detail 24 different setups, summarized in Table \ref{tab:catalogue}. For each set of simulations, we post-processed three different projections (to account for possible biases not seen in observations) at different time steps. The number of saved time steps changes per realization, as each computational time step is associated with a specific time in years and a sink formation efficiency computed, which themselves depend on the initial parameter of each simulations. Defining the star (or sink) formation efficiency (SFE) as the fraction of total sink mass $M_{\mathrm{tot}} = \sum_{\mathrm{i=1}}^{N_{\mathrm{sink}}} M_{\mathrm{i}} $ over the initial gas mass $M_0$, i.e.,
\begin{eqnarray}
    \mathrm{SFE} \equiv \frac{M_{\mathrm{tot}}}{M_0},
\end{eqnarray}
we sampled an evolution range of ${\sim}10^5\,$years from the formation of the first sink particle and we stop the calculations at a SFE of ${\sim}15\%$ and ${\sim}30\%$ for the $500\,M_{\odot}$ and $1000\,M_{\odot}$ realizations, respectively (Paper I). With this in mind, we ended up with a variable number of time steps between 7 and 15 for a total of 732 synthetic observations. For this work we will discuss the results obtained from the post-processing in 6 different wavelengths, from $24\mum$ to $500\mum$, for a total of 4392 maps, as detailed in the following sections (the results from the post-processed maps at 1.3\,mm are discussed in Paper III).

\begin{table}
  \caption{Parameter range of the RS1.0 catalog of simulations analyzed in this work.}     
\label{tab:catalogue}      
\centering          
\begin{tabular}{c c }     
\hline\hline 
Parameters & Values \\
\hline\hline 
Mass [$M_{\odot}$] & [500, 1000] \\
Radius [pc] & [0.38] \\
Mach & [7, 10] \\
$\mu$ & [3, 10, 100] \\
Seed & [1, 2] \\
\hline \hline
\end{tabular}\\
\end{table}

%%%%%%%%%%%%%%%%%%%%%%%%%%%%%%%%%%%%%%%%%%%%%%%%%%%%%%%%%%%%%%%%%%%%%%%%%%%%%%%%%%%%%%%%%%%%%%%%%%%%%%%%%%%%%%%%%%%%%%%%%%%%%%%%%%
\subsection{Multi-wavelength continuum radiative transfer}

\subsubsection{Monte Carlo thermal calculations}

The simulation outputs from \textsc{Ramses} are then post-processed employing 3D Monte Carlo (MC) radiative transfer on the native AMR grid (i.e., with the exact resolution and distribution as the \textsc{Ramses} simulations) using the \textsc{Radmc-3d} code (\citealt{2012ascl.soft02015D}). For the goals of the RS project, the imaging part of the post-processing is done for seven wavelengths. The five Hi-GAL wavelengths (from 70$\mum$ up to 500 $\mum$, \citealt{Molinari10_PASP}) and the additional $24\mum$ to mimic the MIR emission from MIPSGAL \citep{2009PASP..121...76C} are required to properly estimate the bolometric luminosity. In addition, the $1.3\mm$ ALMA Band 6 emission is used to investigate the fragmentation properties at ${\simeq}5000\au$ resolution in comparison with the results of the ALMA-SQUALO survey \citep{2023MNRAS.520.2306T} by means of a dedicated ALMA pipeline as detailed in Paper III.

The post-processing has been done with a similar approach as in \cite{2024A&A...684A..36T}. One important difference is that the dust temperature is re-calculated consistently in \textsc{Radmc-3d} during the post-processing, instead of assuming a thermal equilibrium between the dust and the gas as in \citealt{2024A&A...684A..36T}. The reason is the multi-wavelength nature of this study, as opposed to the single millimeter wavelength analysis of protoplanetary disks in \citet{2024A&A...684A..36T}. In addition, the coarser resolution of the RS1.0 simulations compared to the ones used to simulate the formation of protoplanetary disks in \citet{2024A&A...684A..36T} allow us to carry out the thermal MC calculations without too much computational cost. In particular, we assume a standard dust-to-gas mass ratio of $0.01$, similar to what has been done in Planck and Rosseland mean opacity computations, to infer the dust density profile for \textsc{Radmc-3d} from the gas density calculated with \textsc{Ramses} (Paper I). The gas density provided by the \textsc{Ramses} simulations is then used as an input for both the MC thermal calculation and the imaging process through ray-tracing in \textsc{Radmc-3d}. Then, the properties of the sink particles are read and provided to \textsc{Radmc-3d} as the photon sources with point-like geometry and blackbody spectrum, which is an adequate representation of newly formed protostars with atmospheres not yet developed.

For the energy of the stars, we consider two types of luminosity, the internal luminosity of the sinks, $L_{\rm int}$, computed using the stellar evolutionary tracks from \cite{2013ApJ...772...61K}, and the luminosity resulting from the accretion feedback effect, or accretion luminosity:
\begin{equation}
    L_{\rm{acc}} = f_{\rm{acc}} \frac{\mathcal{G} M_{\rm sink} \dot{M}_{\rm sink}}{R_{\star}}.
\end{equation}
Here $\mathcal{G}$ is the gravitational constant, $\dot{M}_{\rm sink}$ is the mass accretion rate onto the sink particle, and $R_{\star}$ is the stellar radius estimated from the same evolution models used for the calculation of $L_{\rm int}$. The unknown efficiency factor $f_{\rm{acc}}<1$ accounting for how much the gravitational energy is converted into radiation at very small scales is chosen to be $f_{\rm{acc}}=0.1$, in line with what was shown by extremely high-resolution AMR simulations of protostar formation resolving the star-disk connection without the need for sub-grid models \citep{Ahmad2024}.

\begin{figure}
    \centering
    \includegraphics[width=\columnwidth]{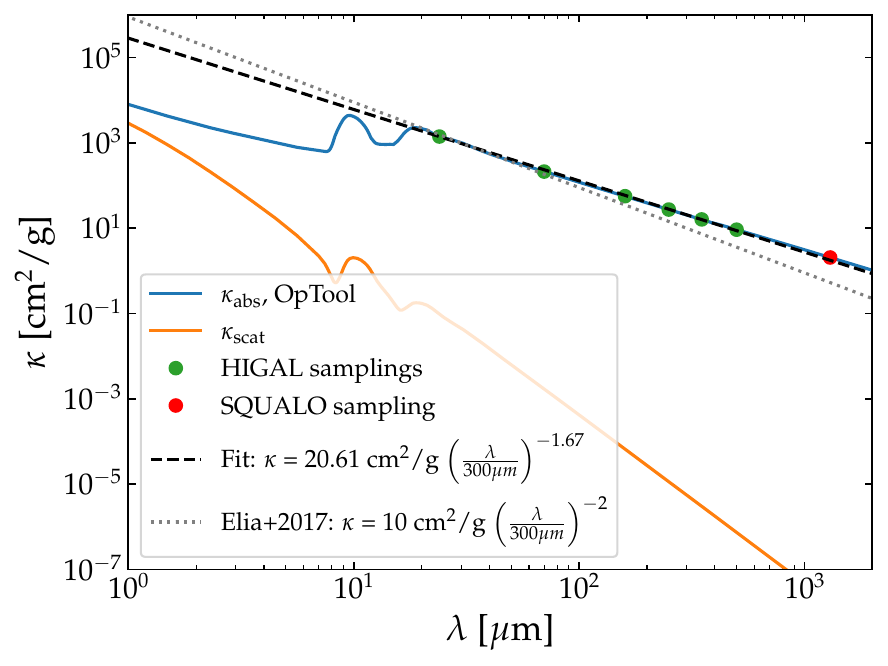}
    \caption{Absorption ($\kappa_{\rm abs}$; blue) and scattering ($\kappa_{\rm scat}$; orange) opacities from \texttt{OpTool} as a function of the wavelength used in this study. For the calculation of the clump masses in Sect. \ref{sec:analysis}, $\kappa_{\rm abs}$ is sampled at Spizer $24\mum$ and Hi-GAL wavelengths (green dots), and at $1.3\mm$ (red dot) for the ALMA observations done in SQUALO, which corresponds to the power-law scaling given by the black dashed line, in comparison with the opacity scaling law used in \cite{2017MNRAS.471..100E} (black dotted line).}
    \label{fig:opacity}
\end{figure}

In the \textsc{Radmc-3d} radiative transfer implementation the optical properties of the dust are encoded in the dust opacity, which is generated here using the tool for dust particle opacities calculations, \textsc{OpTool} (\citealt{2021ascl.soft04010D}). For the RS project we need to overcome the limitations of the (averaged) Planck and Rosseland mean opacities used in \textsc{Ramses}. For this purpose, we adopt the DIANA standard dust model to compute the wavelength-dependent absorption and scattering opacities as required by \textsc{Radmc-3d} for the input opacity file. This model comprises a specific pyroxene ($70\%$ Mg) and carbon, in a mass ratio of $0.87/0.13$, and with a porosity of $25\%$ (\citealt{2016A&A...586A.103W}). These grains are assumed to follow the MRN power-law size distribution with the minimum size $a_{\min} = 0.005\mum$ and the maximum size $a_{\max} = 0.25\mum$ in radius as in the ISM (\citealt{1977ApJ...217..425M}), which are then mass-averaged to determine the absorption and scattering opacities, $\kappa_{\rm abs}$ and $\kappa_{\rm scat}$ respectively, over a wavelength grid from $1~{\rm nm}$ to $1\cm$ comprising 200 wavelength bins. In Fig. \ref{fig:opacity} we show $\kappa_{\rm abs}$ and $\kappa_{\rm scat}$ calculated in the wavelength range $1\leq\lambda\leq 2000\ \mum$.

With these prescriptions, we set up the thermal MC runs with $10^7$ photon packages for each of the sources present in each simulation snapshot to obtain the 3D temperature distribution with \textsc{Radmc-3d}. We chose $10^7$ photon packages after testing also the thermal MC runs with $10^6$ and $10^8$ photon packages. In the former case, the photon packages were not enough to consistently distribute the photons across our simulations, and in the latter, the difference with $10^7$ photon packages was minimal. Both $10^7$ and $10^8$ photon packages reached the convergence, but the computational cost of using $10^8$ photon packages was significantly higher. We have used the default setting \texttt{mc\_weighted\_photons=1} in the code, which distributes the user-specified photon number equally to the sources but assigns different energy to the photon packages (with different `weightings') based on the stellar luminosity. The modified random walk (MRW; \citealt{2009A&A...497..155M}; \citealt{2010A&A...520A..70R}) mode is switched on to speed up the code in regions of high optical depth ($\tau \gtrsim 10$) around the sink particles, in which the photon packages would otherwise risk having to reprocess multiple times or getting stuck for unnecessarily long time before being able to escape. 

\subsubsection{Ray-tracing}

Upon obtaining the dust temperature, we ray-trace through the grid to image the entire clumps, assuming a distance $d=5.2~{\rm kpc}$ to the observer, from three viewing angles that provide the full view of the simulation box in three different planes of the internal simulation grid: $xy$, $xz$, and $yz$. The 2D regularly spaced camera observing the 3D AMR grids comprises $2048$ pixels in each dimension. The pixel size corresponds to a physical resolution of ${\approx}154\au$, which is adequate to produce simulations able to be compared with actual ALMA observations, such as the ones of the SQUALO survey ($5000{-}7000\au$, Paper III) and the newly released ALMAGAL survey (${\sim}1500\au$, \citealt{2025A&A...696A.149M}). By design, the ray-tracing in \textsc{Radmc-3d} is preceded by a MC scattering run to determine the scattering source function that contributes to the final intensity. For this MC scattering run we chose a different set of parameters, compared to the thermal ones: we assign $10^9$ scattering photon packages for all the stars at the Hi-GAL wavelengths and $10^{10}$ photons for the $24\mum$ imaging. The higher number of photons for the Spizer wavelength is based on our pre-testings of the ray-tracing process on various simulation models. These tests show that a total of $10^9$ scattering photon packages for all the photon sources (which is equivalent to ${\sim}10^7$ photon packages for each source in the late-stage snapshots), have been sufficient for the Hi-GAL images, where scattering is negligible. At the same time, they produced a high level of photon noise at 24 $\mum$ due to a combination of scattering effects and lower specific intensity compared to the Hi-GAL wavelengths, which can only be avoided by increasing the number of scattering photons by one order of magnitude. Additionally, we now set \texttt{mc\_weighted\_photons=0} instead of \texttt{1} as in the thermal MC to distribute the photon packages to the sources based on their luminosity, with the more luminous stars getting more photons, to avoid wasting photons on the fainter sources, especially when they are not visible in the final image. The anisotropic scattering using the Henyey-Greenstein formula (\citealt{1941ApJ....93...70H}; mode 2 in the code) is chosen for realistic scattering treatment without the complexity and computation cost of the full polarized scattering mode using the scattering Müller matrix.

We recall that our goal is to infer the $L/M$ parameter from the simulations using the same approach as in observations and to compare it with the evolutionary parameters derived from the simulations, namely the clump's age and, in particular, the SFE. To this end, we have generated realistic synthetic observations in the wavelength range $24 \leq \lambda \leq 500\mum$ and for various time steps of the simulations, corresponding to a clump age range $0$ to ${\sim}0.25$ Myr and on average a SFE range of $[0,0.15]$ for clumps with mass of $500\,M_{\odot}$ and a SFE range of $[0,0.3]$ for clumps with masses of $1000\,M_{\odot}$. For our 10 setups of simulations and the three projections, the final number of snapshots used to produce synthetic simulations is 732 for each wavelength. As discussed in the next sections, we use these synthetic observations to derive the $L/M$ parameter for each time step and compare that with the corresponding clump age and SFE.

%%%%%%%%%%%%%%%%%%%%%%%%%%%%%%%%%%%%%%%%%%%%%%%%%%%%%%%%%%%%%%%%%%%%%%%%%%%%%%%%%%%%%%%%%%%%%%%%%%%%%%%%%%%%%%%%%%%%%%%%%%%%%%%%%%
\subsection{Synthetic Hi-GAL observations and source extraction}
To generate realistic Hi-GAL synthetic observations, we convolved our synthetic maps with the point spread function (PSF) of the $\textit{Herschel}$ PACS \citep[$70$ and $160 \mum$ maps;][]{Poglitsch10} and SPIRE \cite[$250$, $350$ and $500 \mum$ maps;][]{Griffin10} instruments, approximated by a circular 2D Gaussian in each wavelength. The standard deviation of the simulated PSF is equal to $\sigma = {\rm FWHM}/(2\sqrt{2\ln 2})$, where FWHM is the full widths at half maximum of the Hi-GAL beams at the wavelength in consideration, which measure ${\simeq}[8.44, 13.5, 18.2, 24.9, 36.3]''$ for $[70, 160, 250, 350, 500]\mum$ respectively (\citealt{2011MNRAS.416.2932T}). 

\begin{figure*}
    \centering
    \includegraphics[width=\textwidth]{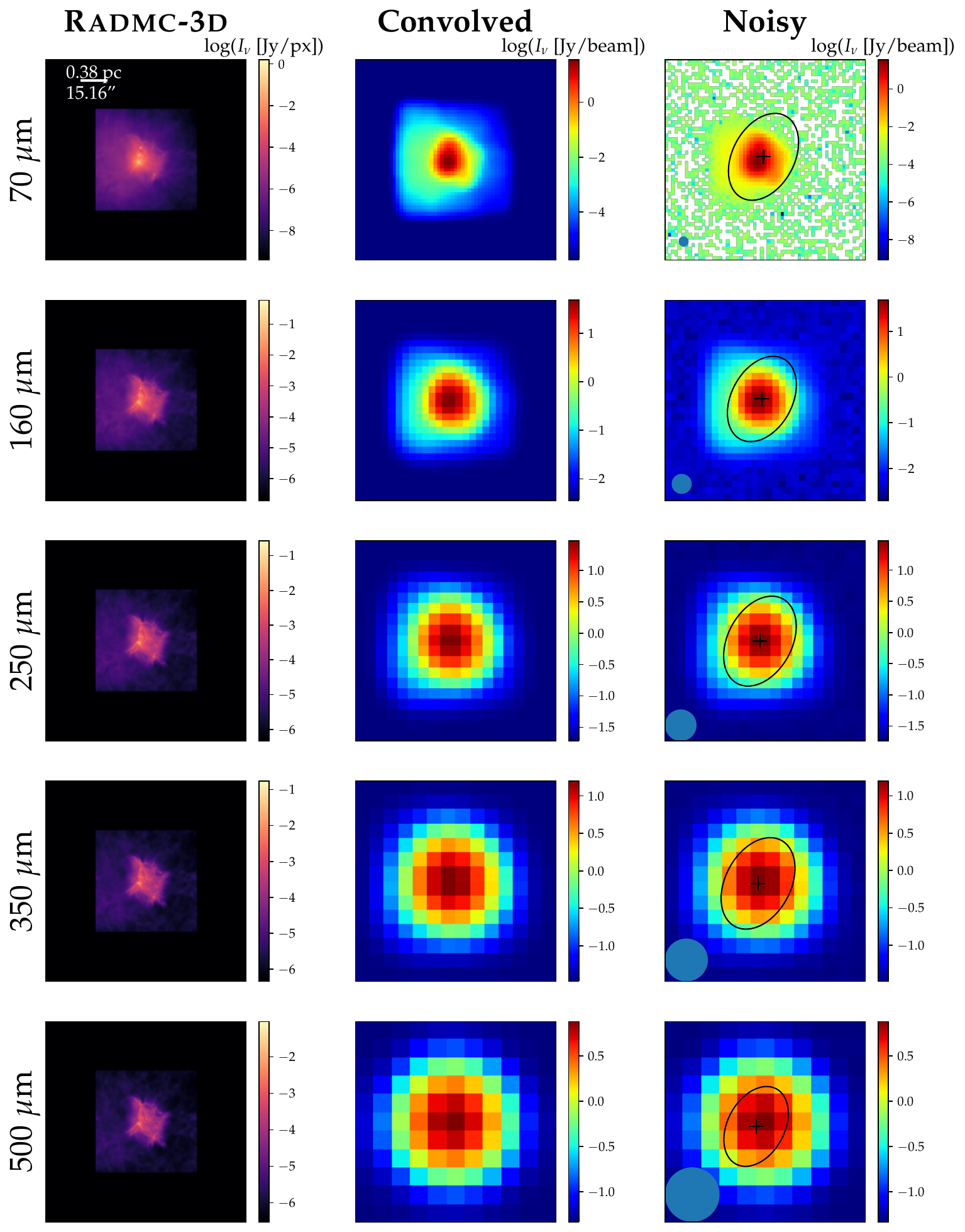}
    \caption{\textsc{Radmc-3d} images (left column) with the Hi-GAL beams convolved (middle column) and the instrumental noise added (right column). The source size extracted by \textsc{Hyper} is the illustrated by the ellipses and the beams by the filled circles.}
    \label{fig:beamconvolution}
\end{figure*}

The instrumental noise in the camera is approximated by the white Gaussian noise from the noise root mean squared, which measures ${\rm rms}=[15.85, 8.01, 3.10, 1.81, 0.94]~{\rm MJy/sr}$ in the Herschel bands and are computed as the median of the mean rms distribution within the Hi-GAL tiles across the Galactic plane reported in \cite{2016A&A...591A.149M} and plotted in Fig. \ref{fig:rms}. For the $24\mum$ wavelength, which is used only for the computations of the bolometric luminosity, we use the Spitzer MIPSGAL survey (\citealt{2009PASP..121...76C}) beam and noise rms, which corresponds to a FWHM of ${\simeq}6''$ and a rms noise of ${\simeq}0.67~{\rm mJy/px}$. Figure \ref{fig:beamconvolution} demonstrates the visual changes of the sources at each step, from the ideal specific intensity map imaged with \textsc{Radmc-3d} from the MC radiative transfer outputs to the noisy maps with the instrumental noise added and the telescope's beams convolved for the Hi-GAL wavelengths.

In order to perform the flux extraction of our clump in the post-processed maps we used \textsc{Hyper},\footnote{The latest Python-based version of the \textsc{Hyper} code is available at \url{https://github.com/Alessio-Traficante/hyper-py}}  software designed to perform aperture photometry in particular on \textit{Herschel} images,  as was done in the Hi-GAL survey (\citealt{2015MNRAS.451.3089T}), with specific versatility on the estimate and subtraction of highly variable background (see \citealt{2015MNRAS.451.3089T} for technical details on the background subtraction). For each of the source detected in the maps \textsc{Hyper} returns, among other parameters, the coordinates of the source peak, the semi-minor, the semi-major axes and the position angle of the 2D-Gaussian fit. Following the approach described in \citet{2015MNRAS.451.3089T}, the 2D-Gaussian estimated at $250\mum$ are used as reference to define the region over which integrate the flux at all wavelengths within the same emitting area. For illustration purposes, the source geometry extracted with \textsc{Hyper} from the post-processed maps of the fiducial model is depicted in the bottom panels of Fig. \ref{fig:beamconvolution}.

These fluxes are used to estimate the mass and luminosity of our clumps with the same approach used in observations in order to obtain a $L/M$ parameter from our synthetic observations comparable with the same value extracted from real observations. We will discuss how to estimate $L$ and $M$ from our extracted fluxes in the next Section.

%%%%%%%%%%%%%%%%%%%%%%%%%%%%%%%%%%%%%%%%%%%%%%%%%%%%%%%%%%%%%%%%%%%%%%%%%%%%%%%%%%%%%%%%%%%%%%%%%%%%%%%%%%%%%%%%%%%%%%%%%%%%%%%%%%
\section{Analysis and results}\label{sec:analysis}

\subsection{Luminosity and temperature reconstruction}\label{sec:LandM} 

The bolometric luminosity $L$ can be obtained from the integration of the area below the spectral energy distribution (SED) curve,

\begin{equation}
    L_{\rm bol} = \int F_{\nu} d\nu,
\end{equation}
for which we use the fluxes at the Hi-GAL wavelengths and the additional $24\mum$ counterpart.

\begin{figure}
    \centering
    \includegraphics[width=0.5\textwidth]{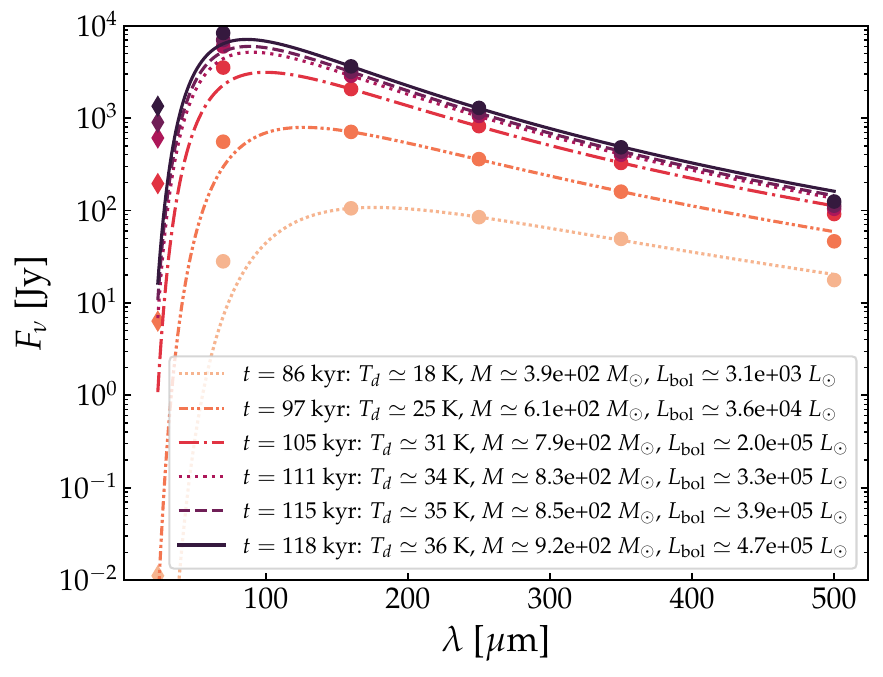}
    \caption{Time evolution of the spectral energy distribution (SED) of the clump shown in Fig. \ref{fig:beamconvolution}. The circle markers show the fluxes of the source as identified by \textsc{Hyper}, whereas the diamond markers indicate the additional fluxes at $24\mum$ used for the bolometric luminosity calculation. The area under this SED is integrated to determine the $L_{\rm bol}$. The optically thin fluxes at $160, 250, 350,$ and $500\mum$ are fitted to a graybody to determine the mass and the average temperature of the clumps. The results of these calculations for this clump are noted in the legend of the figure. }
    \label{fig:sedfit}
\end{figure}

Using a similar approach to \cite{2017MNRAS.471..100E} for the mass and temperature determination, we first write the expression for the total flux $F_{\nu}$ of a graybody at frequency $\nu$ and its approximation under the optically thin assumption that gives the direct relation between the flux and the mass $M$ as (see also \citealt{Pezzuto12})

\begin{eqnarray}
    F_{\nu} &=& (1-e^{-\tau_{\nu}})B_{\nu}(T_d)\Omega\nonumber\\
            &\simeq& \frac{\kappa_{\nu}B_{\nu}(T_d)M}{d^2},
    \label{eq:flux1}
\end{eqnarray}
where $B_{\nu}$ is the Planck function at the dust temperature $T_d$, $\Omega$ the source solid angle in the sky equivalent to the source area extracted by \textsc{Hyper}, and $\kappa_{\nu}$ the absorption opacity, directly linked to the wavelength-dependent optical depth $\tau_{\nu}$ that can be parameterized as
\begin{eqnarray}
    \tau_{\nu} \equiv (\nu/\nu_0)^{\beta} =(\lambda/\lambda_0)^{-\beta},
    \label{eq:tau}
\end{eqnarray} 
with $\nu_0=c/\lambda_0$ being the cut-off frequency at which $\tau_{\nu_0} = 1$, and $\beta$ the exponent of the power-law dust emissivity in the Rayleigh–Jeans regime which in our case is equal to $1.67$ as indicated by the power-law fit for the absorption opacity $\kappa_{\rm abs}$ shown in Fig. \ref{fig:opacity}.

In the optically thin case ($\tau \leq 0.1$ at the minimum wavelength in consideration here, $\lambda=160\mum$), the optical depth is negligible and Eq. (\ref{eq:flux1}) can be simplified as
\begin{eqnarray}
    F_{\nu} = \frac{M\kappa_{\rm ref}}{d^2} \left( \frac{\nu}{\nu_{\rm ref}} \right)^{\beta} B_{\nu}(T_d),
    \label{eq:flux2}
\end{eqnarray}
where the reference wavelength $\lambda_{\rm ref}=300\mum$ is chosen as in \cite{2017MNRAS.471..100E}, which corresponds to $\kappa_{\rm ref}=0.2061\cm^2/\g$ assuming a dust-to-gas mass ratio of $0.01$ (cf. Fig \ref{fig:opacity}). We note that for the sake of consistency, this value of $\kappa_{\rm ref}$ is scaled, according to the opacity of the dust model employed in the \textsc{Radmc-3d} radiative transfer, from what was used in \cite{2017MNRAS.471..100E}, which is $0.1\cm^2/\g$ at $300\mum$. Using this relation, we fit for the two parameters $T_d$ and $M$ from the total fluxes measured by \textsc{Hyper} at four optically thin wavelengths $[160, 250, 350, 500]\mum$. The fittings are done following the same procedure detailed in \cite{2017MNRAS.471..100E}. In essence, Eq. (\ref{eq:tau}) indicates that the critical value of $\tau=0.1$ is encountered at $\lambda_0 \approx 40.3\mum$ for the chosen $\beta=1.67$. Therefore, we first fit for $\lambda_0$ using Eq. (\ref{eq:tau}) by letting its value vary between $5$ and $350\mum$, and Eq. (\ref{eq:flux1}) by letting $T_d$ vary between $5$ and $70\K$. If the value obtained is smaller than $40.3\mum$, we repeat the fit using Eq. (\ref{eq:flux2}) and take the new values of $M$ and $T_d$ as the correct values for the mass and average temperature of the source. This is done for all the time steps considered in each set of realization, except for the very first time step that is produced and in which the presence of sink particles is recorded, owing to the low level of fluxes and the discrepancy between the temperature distributions from the \textsc{Ramses} simulations and the \textsc{Radmc-3d} thermal MC calculations as discussed in Appendix \ref{app:massfit}. For one realization (with seed 2, $\mu=3$, $\mathcal{M}=10$), we also exclude the second time step due to the flat SED from $160$ to $500\mum$ and the very low level of $L/M \sim 10^{-3}\,L_{\odot}/M_{\odot}$ derived from the simulation. An example of the time evolution of the SED for our fiducial model is in Figure \ref{fig:sedfit}. Here we fit the SED in its optically thin regime ($\lambda>160\,\mum$) to estimate the dust temperature and the clump mass. The Figure shows the $70\mum$ and $24\mum$ excess with respect to the SED fitting, as expected when some protostars (i.e., sink) start to illuminate the dust envelope \citep{2015MNRAS.451.3089T, 2017MNRAS.471..100E}.

The results are plotted in Fig. \ref{fig:Mass_lum_time}, as a function of the age of clump since the start of the simulations. The black points are the luminosity recorded in the simulations, which constitutes the sum of the luminosity of all the stars (sink particles) present, and the mass of the gas at each time step, defined as the initial clump mass minus the mass locked into the sink particles. Here, there is a visible detachment of the realizations with $\mu=3$ (the light green points) in comparison with the ones with other values of $\mu=10,100$ which tend to organize themselves into two groups, determined by the initial clump mass. The realizations that started with a clump mass of $1000\,M_{\odot}$ reach high luminosities much sooner than the realizations that started with a clump mass of $500\,M_{\odot}$. This is expected since the radius in both sets of simulations is the same, therefore the more massive clump is also the denser and it collapses more rapidly under the influence of its own gravitational potential (see Paper I). As a result, the sink particles are produced sooner and in larger number and the more massive clump reaches high luminosity sooner then the less massive clump. At the same time, the different parameters of the realizations such as different mach number of different magnetic-to-flux-mass ratio $\mu$ produce a scatter in the distributions, but less pronounced than the differences produced by the different initial mass of the clumps.

Compared to the total luminosity recorded in the simulation the bolometric luminosity is relatively well recovered in each realization and for all time steps (Fig. \ref{fig:Mass_lum_time}). The clump masses, on the other hand, have a more complex behavior: there is a large discrepancy in particular in the very first time steps of the simulations, while the inferred and the simulated clump mass tends to converge towards more advanced time steps.\footnote{We refer hereafter to the mass of the region identified with \textsc{Hyper} as {\it clump mass} and the total gas mass within the simulated region determined in the theoretical analysis as {\it box mass}.}

In Appendix \ref{app:massfit}, we perform several additional experimental fits for the total mass contained within the simulation box to further investigate the accuracy of the mass determination procedure, in particular by estimating the mass using a pixel-by-pixel column density estimation from the original \textsc{Radmc-3d} specific intensity maps, which would take into account possible temperature variations along each pixel, instead of integrating the SED fluxes across the 2D-Gaussian region. We note that this approach gives indeed more reliable results in the simulations. In fact, building the SED by using a single average value of the temperature may be inaccurate for the possible inhomogeneity in the temperature profile, especially in the early time steps of formation \citep[e.g.,][]{Wilcock11}. However, in practice, this approach is not feasible for observations of compact sources such as our clumps. In observational data, the minimum spatial scale that must be considered is set by the instrumental beam size rather than the pixel size. In most cases, observed clumps are only marginally resolved, typically appearing slightly more extended than the beam size. Consequently, a beam-by-beam column density fitting approach would generally encompass one, or at most two, beams per clump. This approach is effectively equivalent to integrating the flux over the 2D-Gaussian region defined in the previous section. Among the available methods, the most robust strategy for mass determination is to construct the SED by integrating the flux within a well-defined region at each wavelength. We emphasize that the primary objective of this work is to replicate the most accurate procedures applicable to real observational data. Therefore, we estimate  $L/M$  from the SEDs by integrating the flux at each wavelength within the region defined by the \textsc{Hyper} aperture photometry.

%%%%%%%%%%%%%%%%%%%%%%%%%%%%%%%%%%%%%%%%%%%%%%%%%%%%%%%%%%%%%%%%%%%%%%%%%%%%%%%%%%%%%%%%%%%%%%%%%%%%%%%%%%%%%%%%%%%%%%%%%%%%%%%%%%
\subsection{$L/M$ vs. time and SFE}
Having determined the bolometric luminosity $L$  and the mass $M$ of the sources with the same approach used in observations, we can now investigate how the $L/M$ estimation compares with the values measured from the simulations, and if we can compare them with evolutionary time steps in the simulations. 

\begin{figure*}
\includegraphics[width=\textwidth]{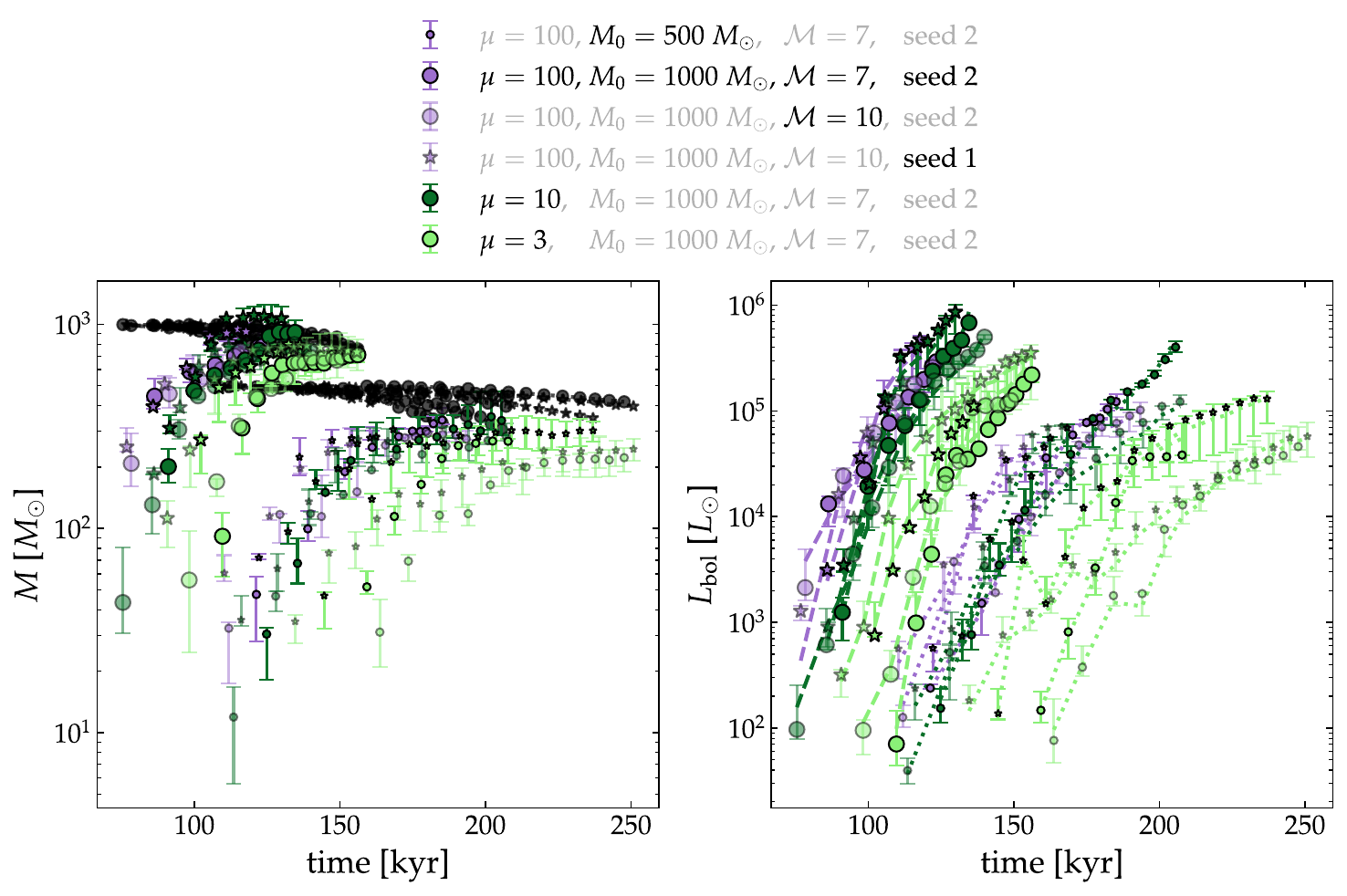}
\caption{Masses of the clumps (left panel) and bolometric luminosities $L_{\rm bol}$ (right panel) extracted with \textsc{Hyper} as a function of the physical time in the simulation. The black points in the left panel mark the closest values inferred from the simulations corresponding to the total gas mass within the clumps minus the mass that ended up in sinks. The total luminosity of all the stars present in the simulation snapshots is plotted by the dashed (for $1000\,M_{\odot}$ initial clumps) and dotted (for $500\,M_{\odot}$ clumps) lines, whose color  coincides with that of the markers for the inferred bolometric luminosity from the synthetic observations. These lines are followed closely by their respective data points, suggesting good agreement between the reconstructed $L_{\rm bol}$ and the total sink luminosities. The error bars indicate the variation of the retrieved values along the three projections considered. Here the change in the colors of the markers indicate the different values of $\mu$, the change in transparency the two values of $\mathcal{M}$ (less transparency for $\mathcal{M}=10$), the change in marker sizes for the initial mass of the clumps (smaller for less massive), and the change in marker types for the seeds of turbulence (stars for seed 1 and circles for seed 2). A representative choice of marker types, sizes, colors, and transparencies are reflect by the parameters in black in the legend, whereas the ones in gray are the parameters of the exact marker they label.}
\label{fig:Mass_lum_time}
\end{figure*}

As discussed in the previous Section, the estimation of the clump mass from the SED fitting approach is not reliable in the first time step, in particular for the temperature structure due to the imprints of the imposed $10\K$ initial value at the beginning of the simulations (see Sec. \ref{sec:sims}). It becomes increasingly reliable at later time steps, achieving an accuracy of approximately $20\%-40\%$ from the second time step onward, and improving significantly at later stages. The estimation of the total luminosity, instead, which is not dependent on the temperature assumption, is very accurate at each time step (see Fig. \ref{fig:Mass_lum_time}). For consistency, in the following analysis we will ignore the results obtained for the first time steps in each realization. It is worth noting that the different parameters of the setups are not affecting this measurement significantly, and the most evident differences between models are due to the initial clump mass.

\begin{figure}
\includegraphics[width=0.5\textwidth]{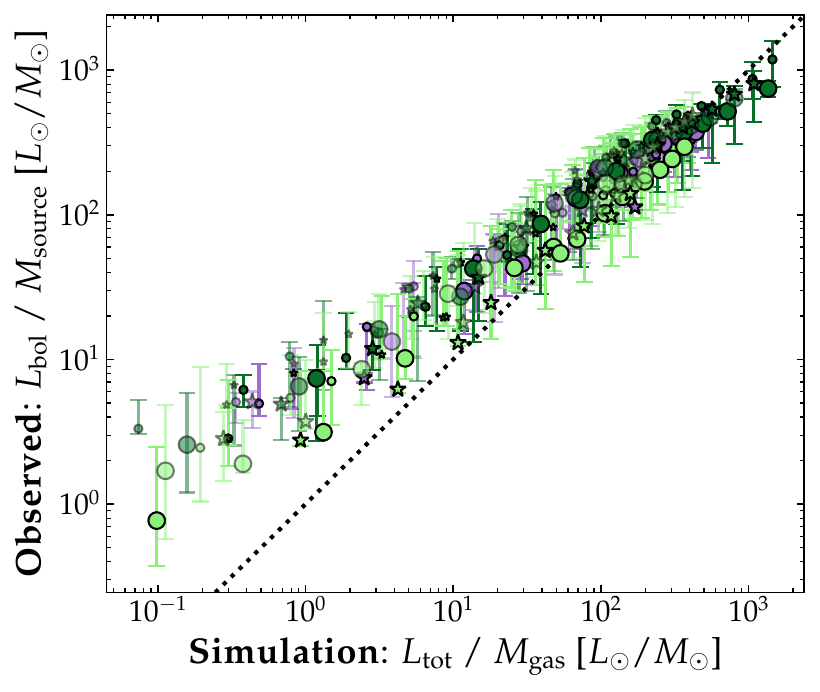}
\caption{$L/M$ computed from our synthetic observations vs. $L/M$ derived from the simulations. As expected from the results shown in Fig. \ref{fig:Mass_lum_time}, in the early time steps the observed $L/M$ is overestimated, as a consequence of the underestimation of the clump mass, and around $L/M\simeq10$ the observed and the simulated $L/M$ starts to converge.}
\label{fig:Lum_obs_vs_lum_theo}
\end{figure}

The value of $L/M$  computed in the simulations using the observational approach is compared in Fig. \ref{fig:Lum_obs_vs_lum_theo} with the theoretical value derived from the simulations, where the luminosity is obtained by summing the total sink luminosity, i.e.,
$L_{\mathrm{tot}} \equiv \sum_{\mathrm{i=1}}^{N_{\mathrm{sink}}}L_{\mathrm{acc,i}}+ \sum_{\mathrm{i=1}}^{N_{\mathrm{sink}}} L_{\mathrm{int,i}},$
and the mass is measured as the total clump mass not locked into sinks, i.e.,
\begin{eqnarray}
    L_{\rm tot}/M_{\rm gas} \equiv \frac{L_{\mathrm{tot}}}{M_0-M_{\mathrm{sinks}}}.
\end{eqnarray}

The plot deviates from the 1:1 relation, as a direct consequence of the differing mass estimation methods used in the simulations and in our observationally driven approach applied to the post-processed maps. This discrepancy has a direct impact on the $L/M$–SFE relation derived from the observational framework, compared to the relation obtained from simulations (Paper I), as further discussed later in this section.

\begin{figure*}
\includegraphics[width=\textwidth]{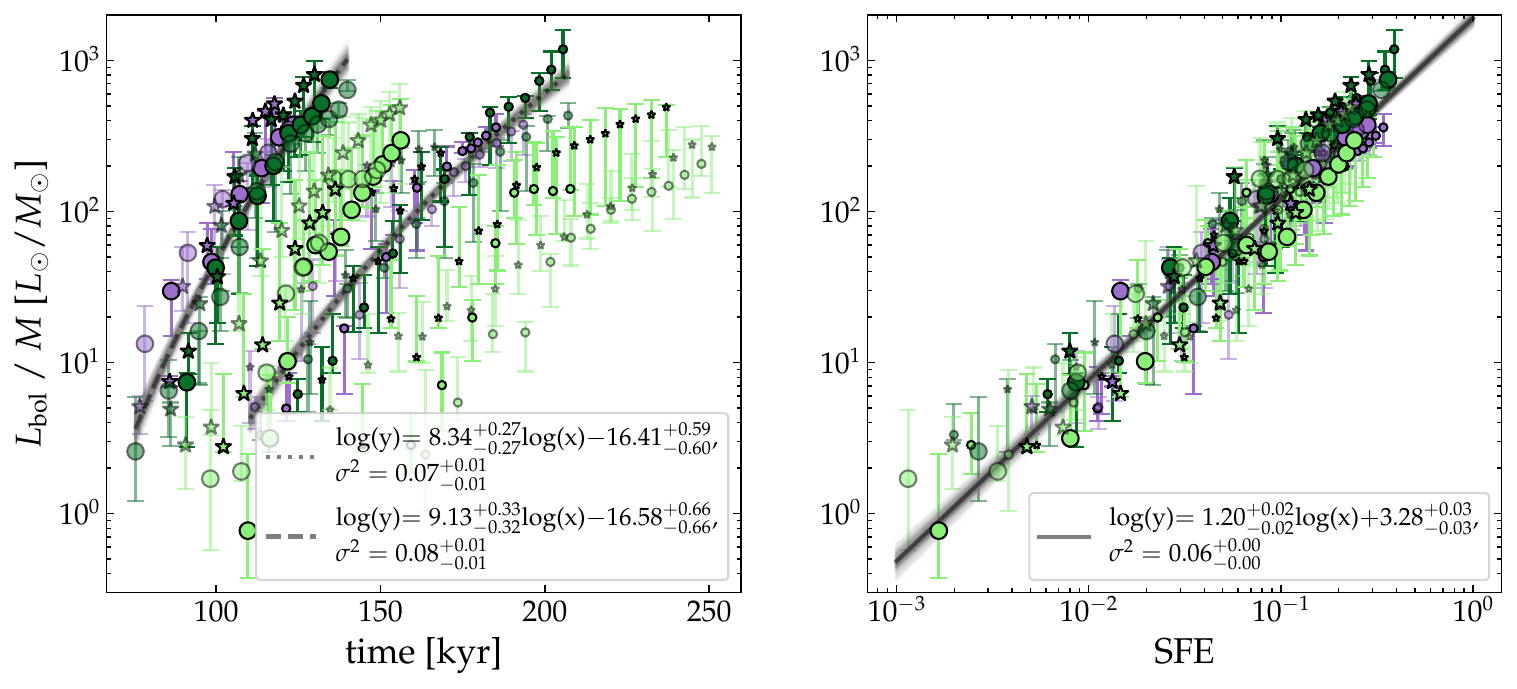}
\caption{\textit{Left panel}: $L/M$ vs. time plot, which shows for each initial clump mass a clear separation between the realizations with $\mu=3$ (in light green) and the ones with other values of $\mu=10,100$ (in dark green and purple, respectively). The latter are then fitted to a log-linear function whose best fits are represented by the back lines for two initial masses. The different setups of each model are color-coded as in Fig. \ref{fig:Mass_lum_time}. \textit{Right panel}: $L/M$ vs. SFE. This relation is independent from the initial clump mass. The black line represents the best fit including all the models described in this work. As for the results in Fig. \ref{fig:Lum_obs_vs_lum_theo}, for small values of $L/M$ the estimates are less accurate due to the uncertainties in the mass estimation, but the theoretical and observed values converge as the clumps evolve.}
\label{fig:L_M_time_SFE}
\end{figure*}

Finally, in Fig. \ref{fig:L_M_time_SFE} we plot the main results of this work. In the left panel we provide  $L/M$  as a function of the physical time since the start of the simulations. Similar to the plots for the luminosity, the points tend to organize themselves into two groups that correspond to the two initial gas mass values explored in the simulations, $M_0 = 500\,M_{\odot}$ and $M_0 = 1000\,M_{\odot}$, and the other parameters do not generate a significant spread compared to one produced by the different initial gas mass, except for the intensity of the magnetic fields, which in the strongly magnetized clumps with $\mu=3$ (light green points) produces a small delay in star formation by a factor of ${\sim}2$. With that in mind, the black curves provide the best log-linear fits to the observationally motivated $L/M$ vs. time relation for the two groups of clumps for the two values of $\mu=10,100$, i.e., in the absence of strong magnetic fields. The fittings are performed with the \texttt{linmix}\footnote{\url{https://github.com/jmeyers314/linmix}} package \citep{2007ApJ...665.1489K} that uses a Bayesian approach for the (log-)linear regression. This approach gives us
\begin{eqnarray}
    \log\left(\frac{L_{\rm bol}}{M}\middle/\frac{L_{\odot}}{M_{\odot}}\right) = 8.34_{-0.27}^{+0.27}\log\left(\frac{t}{1~{\rm kyr}}\right)-16.41_{-0.60}^{+0.59}
\end{eqnarray}
for the $M = 500\,M_{\odot}$ clumps, and
\begin{eqnarray}
    \log\left(\frac{L_{\rm bol}}{M}\middle/\frac{L_{\odot}}{M_{\odot}}\right) = 9.13_{-0.32}^{+0.33}\log\left(\frac{t}{1~{\rm kyr}}\right)-16.58_{-0.66}^{+0.66}
\end{eqnarray}
for the $M = 1000\,M_{\odot}$ clumps. If we compare the results of these fits with an evolutionary classification dividing clumps in the three phases characterized by $L/M\leq1\,L_{\odot}/M_{\odot}$, $1<L/M\leq10\,L_{\odot}/M_{\odot}$ and $L/M>10\,L_{\odot}/M_{\odot}$ of \cite{2016ApJ...826L...8M}, a $500~M_{\odot}$ clump will reach the end of the first phase ($L/M\sim1\,L_{\odot}/M_{\odot}$) at ${\approx}92.8\kyr$ and a $1000\,M_{\odot}$ one at ${\approx}65.5\kyr$. They will then spend ${\approx}29.5$ and ${\approx}18.8\kyr$, respectively, in the second phase (between $1<L/M<10\,L_{\odot}/M_{\odot}$) before entering the third one ($L/M\gtrsim10\,L_{\odot}/M_{\odot}$), the phase in which in some of the most massive clumps there is the possibility of forming \textsc{Hii} regions (which are not yet considered in our current setup for the RS simulations, see Paper I). The absolute age of a clump, however, is likely to be an ill-defined quantity. In the RS simulations we take as age "zero" the moment when we the simulation begins, i.e., when a symmetric sphere of uniform density starts to collapse in an isolated box, but this is not a realistic scenario. Clumps do not form in isolation and the process is a continuous flow of gas from large down to small scales (\citealt{2006ApJ...648.1052H, 2008A&A...486L..43H, 2009MNRAS.398.1082B, 2012MNRAS.424.2599C}).

Instead, the SFE is a much more robust quantity to analyze: the ratio of the mass of the gas locked in stars to the mass observed in a defined clump is a well-defined ratio in simulations: the SFE is 0\% until the first sink forms in a clump, and with time increases until the simulation stops. Furthermore, as shown in Paper I. the SFE as a function of luminosity of the sinks over mass of the clumps is not mass-dependent, as it is the absolute clump age. This is therefore the most robust evolutionary indicator in the simulations to compare with the observed $L/M$ for any given clump. 

As shown in Fig. \ref{fig:L_M_time_SFE}, right panel all the tracks overlap each other in the $L/M$ vs. SFE plot, with the scatter driven by the different initial conditions of the simulations. There a clear correlation between these two quantities and if we fit a power-law function, we obtain the following log-linear relation:

\begin{eqnarray}\label{eq:SFE_L_M}
    \log\left(\frac{L_{\rm bol}}{M}\middle/\frac{L_{\odot}}{M_{\odot}}\right) = 1.20_{-0.02}^{+0.02}\log\left(\mathrm{SFE}\right)+3.28_{-0.03}^{+0.03}\;.
\end{eqnarray}

The best-fit power-law function traces quite closely the data points within less than an order of magnitude of uncertainties among the different setups (in other words, different values of the initial clump mass, turbulent field or intensity of the magnetic fields do not substantially contribute to the $L/M$ vs. SFE relation). The end of the first phase of clumps formation, $L/M=1\,L_{\odot}/M_{\odot}$, is reached at the very beginning of the sink formation: as soon as the first stars appear and the SFE reaches $10^{-3} {-} 10^{-2}$, they start to irradiate the clump and thus its luminosity overcomes the $1\,L_{\odot}/M_{\odot}$ value. The second phase survives until the efficiency grows up to SFE $\simeq10^{-2} {-} 0.5\times10^{-1}$, depending from the parameters of the setup. This result implies that up to $L/M=10$ the vast majority of the star formation activity did not yet take place in the clump. It is warmed up by the first few stars formed in the system. The third phase, clumps with $L/M>10\,L_{\odot}/M_{\odot}$, is when the majority of the star formation occurs.

A noteworthy result is that the $L/M$–SFE relation found in Paper I remains a mass-independent power law, but with a steeper slope of ${\simeq}1.5$. This difference arises from the discrepancy between the $L/M$ values measured in our post-processed maps and those estimated in Paper I, as illustrated in Fig. \ref{fig:Lum_obs_vs_lum_theo}. 

%%%%%%%%%%%%%%%%%%%%%%%%%%%%%%%%%%%%%%%%%%%%%%%%%%%%%%%%%%%%%%%%%%%%%%%%%%%%%%%%%%%%%%%%%%%%%%%%%%%%%%%%%%%%%%%%%%%%%%%%%%%%%%%%%%
\section{Conclusions}\label{sec:conclusions}
In this work, the second in the three-paper series presenting the Rosetta Stone project, we  carried out the post-processing of a set of dedicated simulations (Paper I), produced to explore a wide parameter range of initial clump conditions. Within the RS framework we have post-processed seven wavelengths: Spizer's $24\mum$, five \textit{Herschel} Hi-GAL bands, from $70$ to $500\mum$, and the $1.3\mm$ ALMA Band 6 used to reproduce the SQUALO survey (Paper III). The main results of the current work are as follows:

\begin{itemize}
    \item We presented the post-process of 732 maps of protostellar collapse simulations of clouds that form massive clumps with multi-wavelength 3D MC radiative transfer and produced synthetic observation maps at six wavelengths, from $24\mum$ to $500\mum$. From these maps we extracted synthetic clumps as were done on actual Hi-GAL maps with the aperture photometry code \textsc{Hyper} and analyzed their properties. In particular, we  studied the (observed) bolometric luminosity-to-mass ratio $L/M$ as an evolutionary indicator of massive clumps. 
    \item The clump masses and temperatures determined from the graybody fittings of the SEDs using the isothermal assumption with constant average dust temperature usually used in observational modeling are reliable within $20-40\%$ uncertainty range compared to the realistic values derived from the simulations, in particular for the more evolved stages of the clump formation. The reason for the discrepancies with the models in the youngest phases of formation are likely due to the inhomogeneity of the temperature field that cannot be taken into account with a single-temperature graybody  model.
    \item The bolometric luminosity calculations from the SED yield reliable results in comparison with the total stellar luminosity in the simulation.
    \item In the weakly or moderately magnetized clumps, the correlation between  $L/M$  and the age of the clump is mass-dependent. We found a power-law function of $L/M\propto t^{8.34^{_{-0.27}^{+0.27}}}$ for the $500\,M_{\odot}$ clumps and $L/M\propto t^{9.13^{_{-0.32}^{+0.33}}}$ for the $1000\,M_{\odot}$ clumps. The other parameters explored in our setups either have no significant impact on these correlations (in the case of the initial level of turbulence and the random orientation of the clumps) or potentially a small effect on star formation in the presence of strong magnetic fields, in which scenario a delay by a factor of ${\sim}2$ in the timing of star formation for $\mu=3$ is expected.
    \item The correlation between $L/M$ and SFE, a more reliable indicator of the clump evolution in simulations, is independent from all the initial parameters including the initial clump mass and the intensity of the magnetic field. We found a power-law relation of the form $L/M\propto \mathrm{SFE}^{1.20^{_{-0.02}^{+0.02}}}$ with a little scatter due to the different initial conditions of the setups.
\end{itemize}

We conclude by emphasizing that this study is part of the end-to-end Rosetta Stone framework, and its results are closely linked with those presented in the companion papers. In Paper I, we introduce the Rosetta Stone simulation suite (RS1.0); in the present paper, we detail the post-processing methodology used to generate realistic synthetic maps at various wavelengths from those simulations; and in Paper III, we simulate ALMA observations of the maps produced here to closely replicate the observations from the ALMA-SQUALO survey. 

In particular, our results indicate that $L/M$, as measured in observations, is a reliable tracer of the overall evolutionary stage of massive star-forming regions. While accurately determining the age of observed clumps from this ratio requires a well-constrained estimate of the mass and intensity of the local magnetic fields, $L/M$ can be directly compared with the star formation efficiency (SFE) derived from simulations. Our analysis further shows that the majority of star formation activity, as traced by the SFE, takes place after clumps enter the $L/M > 10$ evolutionary phase.

Further analysis is required to account for additional effects, such as outflows as a source of feedback and the formation of \textsc{Hii} regions. These processes are the central focus of the next series of simulations currently being developed within the RS2.0 framework.

\begin{acknowledgements}
We are grateful to the anonymous referee for insightful comments. This project was funded by the European Research Council via the ERC Synergy Grant ``ECOGAL'' (project ID 855130). We thank the whole consortium for the stimulating discussions which helped us tremendously through this process. This project has received funding from the European Union's Horizon 2020 research and innovation programme under the Marie Sklodowska-Curie grant agreement No 823823 (DUSTBUSTERS). AT gratefully acknowledges support from a mini-grant funded
by INAF. This work was partly supported by the Italian Ministero dell Istruzione, Universit\`a e Ricerca through the grant Progetti Premiali 2012 – iALMA (CUP C$52$I$13000140001$). In addition to the ERC, RSK acknowledges financial support from the German Excellence Strategy via the Heidelberg Cluster ``STRUCTURES'' (EXC 2181 - 390900948) and from the German Ministry for Economic Affairs and Climate Action in project ``MAINN'' (funding ID 50OO2206).  RSK also thanks the 2024/25 Class of Radcliffe Fellows for highly interesting and stimulating discussions. G.A.F gratefully acknowledges the Deutsche Forschungsgemeinschaft (DFG) for funding through SFB 1601 ``Habitats of massive stars across cosmic time’' (sub-project B1) and support from the University of Cologne and its Global Faculty programme.
\end{acknowledgements}

\bibliographystyle{aa_url}
\bibliography{bibliography}

\begin{appendix}

\section{Herschel noise rms}\label{app:rms}

\begin{figure}[ht!]
\centering 
\includegraphics[width=0.45\textwidth]{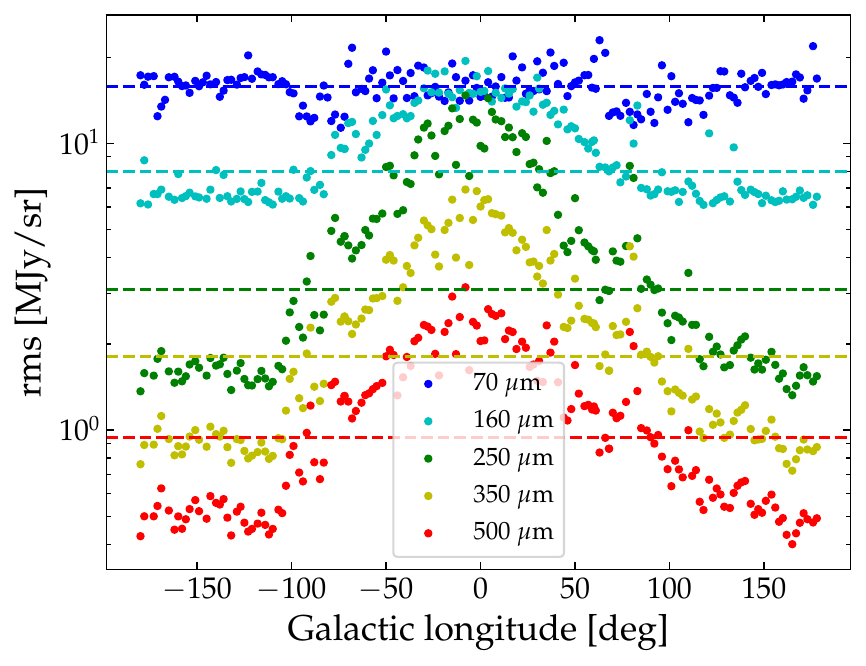}
\caption{Noise rms for each Hi-GAL band averaged across all galactic longitudes from the values reported in \cite{2016A&A...591A.149M}.}
\label{fig:rms}
\end{figure}

The instrumental noise used to generate Hi-GAL synthetic observations are derived from the noise rms for five \textit{Herschel} bands across all galactic longitudes reported in \cite{2016A&A...591A.149M}. In order to construct the random white noise to be added to the ideal specific intensity distribution from \textsc{Radmc-3d} after convolution with the corresponding Hi-GAL beam, we take the median values of the noise rms for each band, as shown by the dashed lines in Fig. \ref{fig:rms}.

\section{Clump mass fittings}\label{app:massfit}

Due to the somewhat expected difference between the clump mass and the total gas mass contained in the simulation box discussed in the main text, we perform further analyses of the gas masses from the Hi-GAL synthetic observations of a specific simulation realization as a benchmark for the mass determination methodology.

\begin{figure*}
    \centering
    \includegraphics[width=\textwidth]{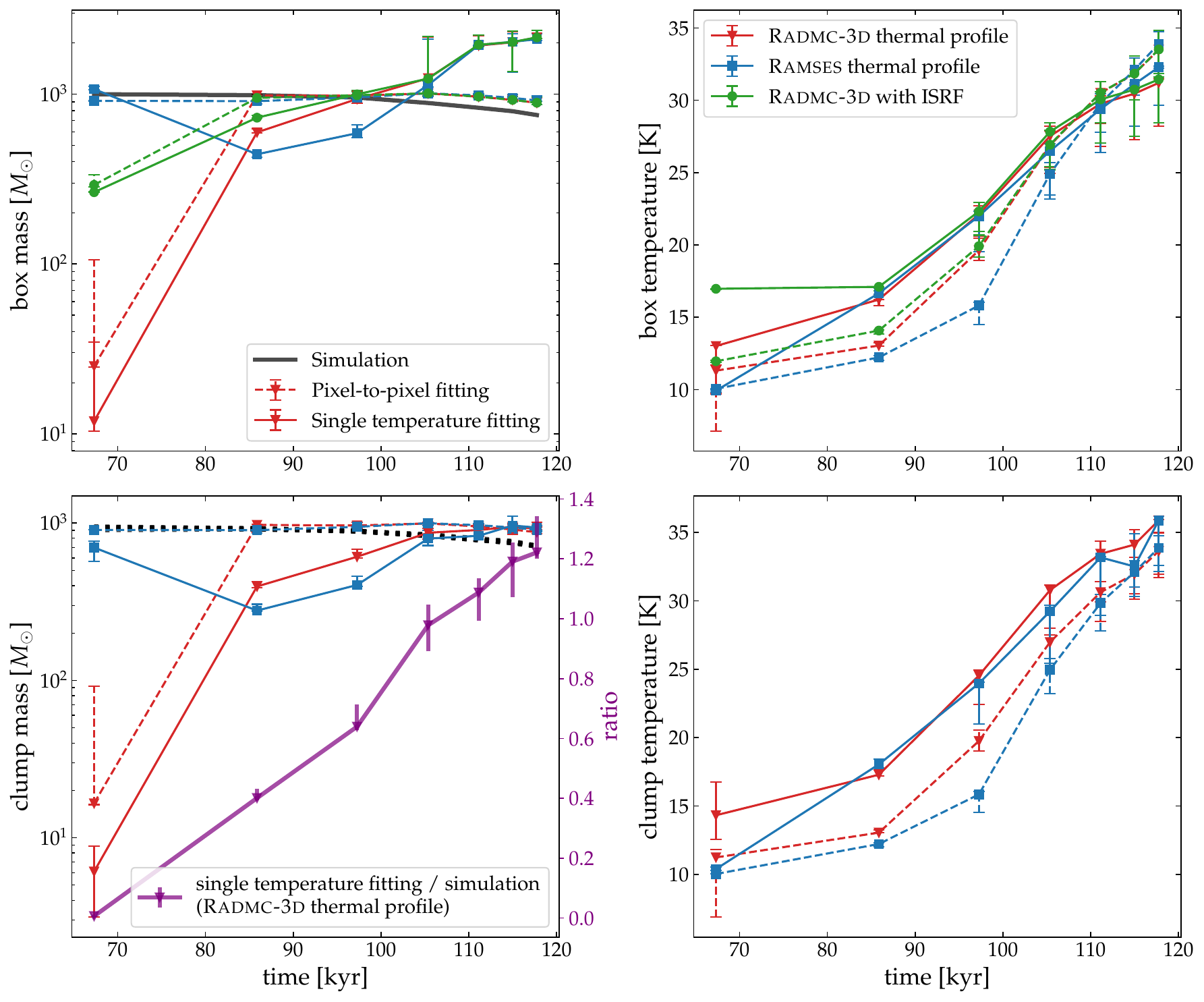}
    \caption{Results of the experimental fits to determine the temperatures and masses of the box (upper panels) and of the clumps (lower panels). Three prescriptions for the dust thermal profile are investigated: full MC radiative transfer calculations (in red), thermal equilibrium between the dust and the gas components (in blue), MC radiative transfer with external radiation source (ISRF; in green). For the box mass and temperature determinations, two approaches are employed: assuming an average constant temperature for the entire box (solid lines) and pixel-by-pixel fittings for column density reconstruction (dashed lines).}
    \label{fig:fitMcld}
\end{figure*}

First, in order to retrieve a mass closer to what is inferred from the total gas mass in the simulation, we use the same approach as the fittings done in Sect. \ref{sec:LandM} to determine the gas mass contained within the region defined by the box size (hereafter referred to as box mass, $M_{\rm box}$) by replacing the source area extracted by \textsc{Hyper} with the entire simulation box area for the value of the solid angle $\Omega$ in Eq. (\ref{eq:flux1}). The fluxes used for the fits are then the fluxes within the entire computational domain. The fitted box mass and average temperature along its evolutionary stages are depicted by the solid red curves in the upper left and right panels of Fig. \ref{fig:fitMcld}, respectively. The same trend as for the clump mass, i.e., the mass increases with time as the flux increases, is observed here due to the increase of the average dust temperature from $T_d=5\K$ to $70\K$ inferred from the blackbody fittings. Most notably, the box mass at the earliest snapshot is more than an order of magnitude lower than the value reported in the simulation (given by the black curve).

Next, to test whether the temperature distribution resulted from the thermal MC calculations plays any role in the mass determination from the observed fluxes, we used two alternative prescriptions for the temperature in the MC radiative transfer with \textsc{Radmc-3d}, namely equating the dust temperature to the gas counterpart available from the \textsc{Ramses} simulation output and imposing an external heating source from the solar neighborhood interstellar radiation field (ISRF). The former approach (i.e., with the thermal profile taken from \textsc{Ramses}), the results of which are plotted by the blue curves in Fig. \ref{fig:fitMcld}, helps alleviate the difference in the retrieved mass at the first snapshot; however the decrease from the second time step and the later convergence with the results from the other two approaches suggest that the MC temperature results start to stabilize from the second output where the imprints of the imposed initial gas temperature of $10\K$ is less significant.

For the latter, the isotropic ISRF with spectrum from \cite{1983A&A...128..212M} is prescribed in \textsc{Radmc-3d} as an external heating source that launches photon packages from outside the simulation box inwards. The inclusion of such radiation field, the results of which are shown by the green curves, on the other hand, overheats the clump, which in reality should be shielded by the parent cloud, and to a greater extent its outskirt to an average temperature above what is seen in the simulation while does not provide an accurate estimate of the box mass in the first time step due to the different physics. We therefore exclude the ISRF in our other calculations, also since it likely provides additional heating that is incompatible with the simulation setup, which does not have the ISRF as a physical ingredient.

\begin{figure*}
    \centering
    \includegraphics[width=\textwidth]{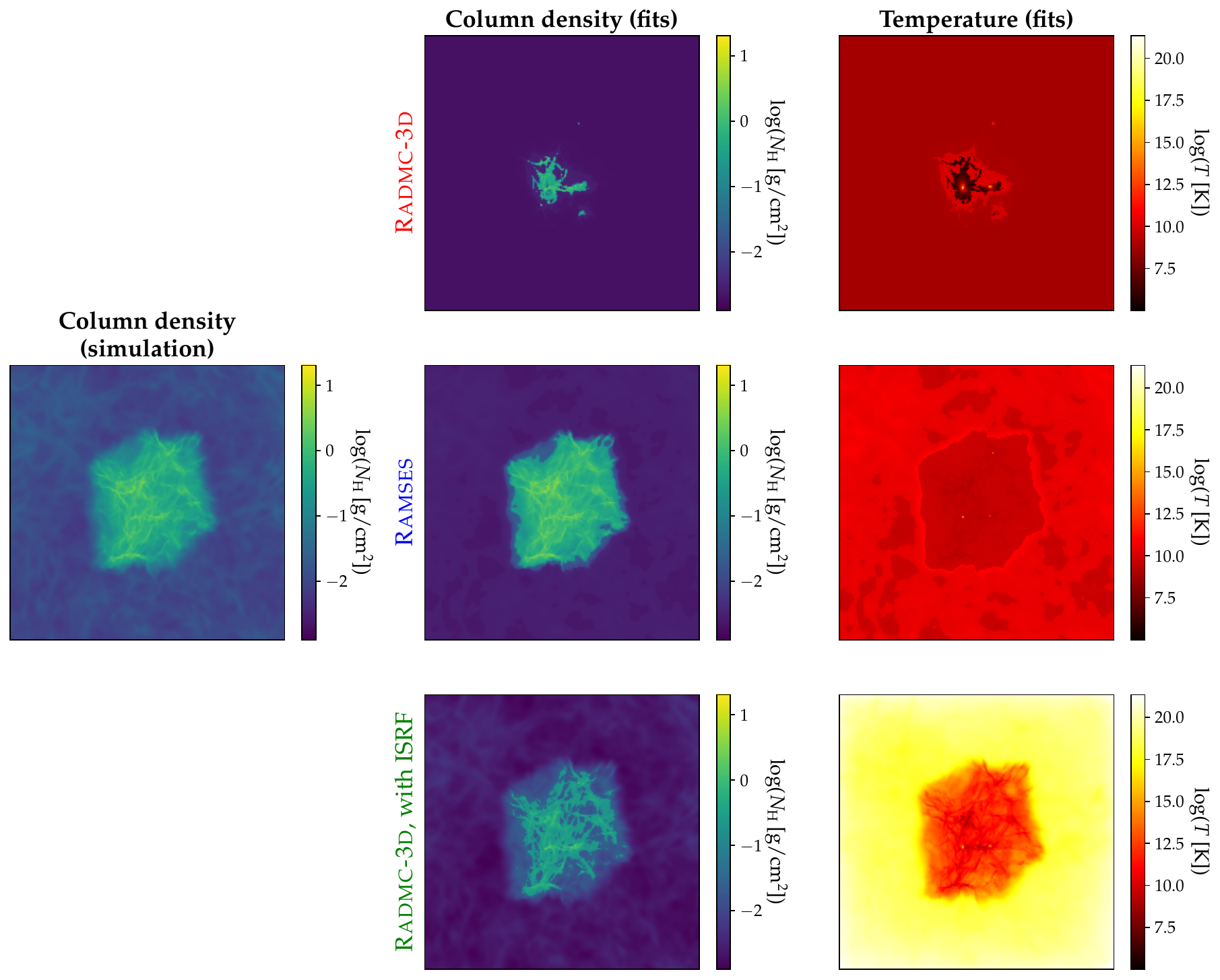}
    \caption{Pixel-by-pixel fitting results for the temperature (right column) and column density (middle column) in the first very output of the fiducial model, in comparison with the original column density from the simulation (left column). Three temperature distributions of the dust are considered for the ray-tracing to obtain the dust emission: one obtained with \textsc{Radmc-3d} thermal MC computations (top row) with stars as the only photon sources, one taken directly from the gas temperature in the \textsc{Ramses} simulation (middle row), and one with the \textsc{Radmc-3d} thermal MC using the ISRF as an additional heating source (bottom row). The \textsc{Radmc-3d} profile without ISRF produces insufficient heating inside the clumps in comparison with the imposed $10\K$ temperature imposed at the beginning of the simulation, resulting in the low fluxes and consequently low mass reproduced. The inclusion of the solar neighborhood ISRF in the \textsc{Radmc-3d} MC treatment, on the other hand, overheats the outskirts of the clump to a much higher temperature than the inside, while still not resulting in an adequate estimate of the mass contained within the latter.}
    \label{fig:pixbypix}
\end{figure*}

	Finally, as a zero-degree analysis to investigate what would be the closest we could get to the "ground truth" with our fitting techniques, we test a more advanced method of fitting pixel-by-pixel the intensity distribution from the ideal emission maps of the clumps. This allows us to build the corresponding temperature and column density maps, as shown in Fig. \ref{fig:pixbypix}, which can then be summed up to obtain the total mass contained within the field of view. In this case, the source solid angle $\Omega$ in Eq. (\ref{eq:flux1}) is replaced by the pixel angular size in steradians. The results with this approach are shown in Fig. \ref{fig:fitMcld} by the dashed line using both the \textsc{Ramses} (in blue) and the MC thermal profile re-calculated with \textsc{Radmc-3d} (in red). Interestingly, for the very first time step, the results using the dust thermal profile from the \textsc{Radmc-3d} MC radiative transfer (the red points) predict poorly the gas mass due to insufficient heating from the first-born star(s), which is not the case when the thermal profile from the \textsc{Ramses} simulation is imposed on the subsequent ray-tracing flux calculation (the blue points). In the same vein, the masses at later stages are overestimated by a factor of ${\simeq}2$ due to the inhomogeneity caused by the much higher temperature inside the clumps compared to the outskirt. That being said, there are significant improvements in the mass obtained here. The fact that the box mass retrieved with the two thermal profiles of the dust agree from the second time step strongly supports the argument that the radiative transfer results in the first step suffer from the uncertainty caused by the temperature initialization in the simulation. If such a step is not taken into consideration, we can assess that the masses of the extracted source (lower left panel) obtained with the single average temperature fittings are reliable with a relative uncertainty of ${\simeq}60\%$ from the next time step, then reduce to ${\simeq}40\%$ before converging to a value that overshoots the actual gas mass inferred from the simulation by ${\simeq}20\%$ (the purple line in Fig. \ref{fig:pixbypix}). This overestimation of the mass at the later, more stable stages, however, is not observed for the less massive clumps (cf. Fig. \ref{fig:Mass_lum_time}), in which case the mass instead tends to be underestimated.

\end{appendix}

\end{document}